\documentclass[twocolumn,aps,prl,showpacs]{revtex4}
\input epsf

\def \snn   {\sqrt{s_{_{\rm NN}}}}
\def \ks    {K_S^0}
\def \pt    {p_\perp}
\def \mt    {m_\perp}
\def \dnhm  {dN_{{\rm h}^-}/d\eta}
\def \dedx  {{\sl dE/dx}}
\def \kink  {{\sl Kink}}
\def \dndy  {{dN/d{\rm y}}}
\def \dndypi {{dN_{\pi^-}/d{\rm y}}}

\def \mdedx {\langle dE/dx \rangle}
\def \lsim  {$\,^{<}_{\sim}\,$}
\def \gsim  {$\,^{>}_{\sim}\,$}

\begin{document}

%%%%%%%%%%%%%%%%%%%%%%%%%%%%%%%%%%%%%%%%%%%%%%%%%%%%%%%%%%%%%%%%%%%%%%

\title{Kaon Production and Kaon to Pion Ratio in Au+Au Collisions at
$\snn=130$~GeV}

\author{
C.~Adler$^{11}$, Z.~Ahammed$^{23}$, C.~Allgower$^{12}$, J.~Amonett$^{14}$,
B.D.~Anderson$^{14}$, M.~Anderson$^5$, G.S.~Averichev$^{9}$, 
J.~Balewski$^{12}$, O.~Barannikova$^{9,23}$, L.S.~Barnby$^{14}$
% ...
\\
(STAR Collaboration)
}
% ...
\address{$^5$University of California, Davis, California 95616}
% ...
\address{$^{9}$Laboratory for High Energy (JINR), Dubna, Russia}
% ...
\address{$^{11}$University of Frankfurt, Frankfurt, Germany}
\address{$^{12}$Indiana University, Bloomington, Indiana 47408}
% ...
\address{$^{14}$Kent State University, Kent, Ohio 44242}
% ...
\address{$^{23}$Purdue University, West Lafayette, Indiana 47907}
% ...

\begin{abstract}
Mid-rapidity transverse mass spectra and multiplicity densities of
charged and neutral kaons are reported for Au+Au collisions at
$\snn$=130~GeV at RHIC. The spectra are exponential in transverse
mass, with an inverse slope of about 280~MeV in central collisions.
The multiplicity densities for these particles scale with the negative 
hadron pseudo-rapidity density. The charged kaon to pion ratios are
$K^+/\pi^- = 0.161 \pm 0.002 {\rm (stat)} \pm 0.024 {\rm (syst)}$ and
$K^-/\pi^- = 0.146 \pm 0.002 {\rm (stat)} \pm 0.022 {\rm (syst)}$ for
the most central collisions.
The $K^+/\pi^-$ ratio is lower than the same ratio observed at the SPS
while the $K^-/\pi^-$ is higher than the SPS result. Both ratios are
enhanced by about 50\% relative to p+p and $\overline{\rm p}$+p
collision data at similar energies.
\end{abstract}

\maketitle

%%%%%%%%%%%%%%%%%%%%%%%%%%%%%%%%%%%%%%%%%%%%%%%%%%%%%%%%%%%%%%%%%%%%%%

Lattice QCD predicts that at sufficiently high energy density, matter
will be in a state of deconfined quarks and gluons~\cite{Laermann}.
It has been suggested that strangeness production is a sensitive probe
of a deconfined state: for example, strangeness production may be
enhanced by the fast and energetically favorable process of 
gluon-gluon fusion into strange quark-antiquark pairs~\cite{Rafelski}.
Hadronic mechanisms, on the other hand, may also enhance strangeness
production~\cite{Sorge}. A systematic investigation of strangeness
production is therefore needed to understand different production
mechanisms.

Strangeness production has been studied in heavy-ion collisions at
the AGS~\cite{E802,E866,E917}, SPS~\cite{NA35,NA49,NA44},
and more recently at RHIC~\cite{phiPaper,PhenixPaper,lambdaPaper}.
In this letter, we report measurements by the STAR experiment at RHIC
on charged and neutral kaon production. The measurements were made at
mid-rapidity in Au+Au collisions at a nucleon-nucleon center-of-mass
energy of $\snn$=130~GeV. The measurements were carried out during
the summer of 2000 and details of the STAR experiment are described
elsewhere~\cite{star,STARhminus,pbarPaper}. The primary tracking
device in the experiment is a Time Projection Chamber (TPC). It sits
in a magnetic field of 0.25 Tesla. Tracks were reconstructed from
three-dimensional hits measured in the TPC and the primary vertex of
the interaction was found by fitting the reconstructed tracks to a
common point of orgin. Corrections were made for the energy loss of
the charged kaons in the detector material. The momentum resolution
was found to have negligible effect on the kaon spectra and so no
correction was applied.

\begin{figure*}[hbt]
\centerline{\epsfysize=0.38\textwidth\epsfbox[0 110 640 410]{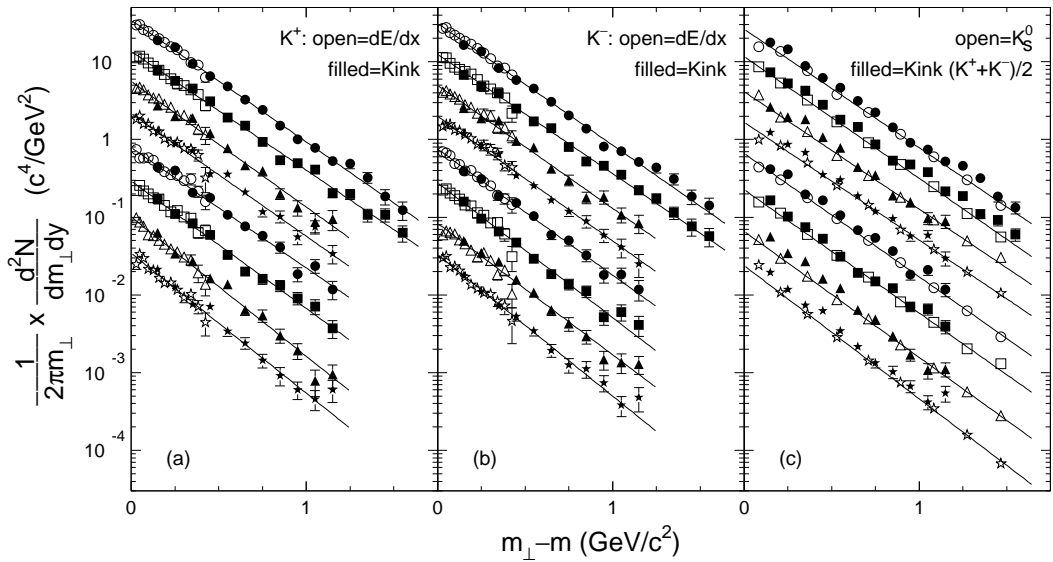}}
\caption{Invariant yields for $K^+$, $K^-$ and $\ks$ versus $\mt$. In
figure (a) $K^+$ and (b) $K^-$, the kaons are identified by the
\dedx\ method ($|y|<0.1$) and by the \kink\ method ($|y|<0.5$). In
figure (c) the $\ks$ is identified by the $\ks$ method ($|y|<0.5$)
and the $(K^+ + K^-)/2$ is from the \kink\ analysis. The data
are plotted in order of decreasing centrality from top to bottom. See
Table~\ref{tab}. The most central collision spectrum is shown full
scale; the other spectra are divided by 2,4,8,16,32,64, and 64 for
display purposes. The error bars are uncorrelated random errors. See
text for a description of the systematic errors. The solid lines are
$\mt$ exponential fits to the $K^+$, $K^-$, and $\ks$ spectra,
respectively.}
\label{spectra}
\end{figure*}

Two methods were used to identify the kaons:

(I) The Energy loss method (\dedx): Particle identification was done
by measuring the mean specific energy lost by the charged particles,
$\langle dE/dx \rangle$, in the TPC gas. The $\langle dE/dx \rangle$
resolution was approximately 11\%. 
The tracks were required to come from within 3~cm of the
primary vertex and every track had at least 25 hits, out of 45
possible hits, on the TPC pad plane. Using a method that is described
in~\cite{pbarPaper}, the distribution in $\ln[\mdedx/\mdedx_{\rm BB}]$
(where $\mdedx_{\rm BB}$ is the expected Bethe-Bloch value) was fit by
a sum of four Gaussians corresponding to $\pi^{\pm}$, $K^{\pm}$,
$e^{\pm}$, and $p$ ($\overline{p}$). The fit was done for each
centrality and transverse momentum ($\pt$) bin. The raw kaon yield
was extracted from the fit parameters. In the range where the kaons
are well separated from other species, $\pt$\lsim0.5~GeV/$c$, we
estimate a point-to-point systematic error of 5\% on the extracted
kaon yields. In the range where the kaons and the $e^{\pm}$ overlap
in $\mdedx$, 0.5\lsim$\pt$\lsim0.7~GeV/$c$, we parameterized the
$e^{\pm}$ yield using data from lower $\pt$ and Monte Carlo (MC) 
simulations and estimate the systematic errors to range from 10\% to 20\%. 
In the region where the kaons significantly 
merge with the pions in $\mdedx$, 0.7\lsim$\pt$\lsim0.8~GeV/$c$, 
we neglect the $e^{\pm}$ contribution and estimate the systematic errors 
to be on the order of 15\%~\cite{AlexThesis}.

(II) The Decay topology method (\kink\ and $\ks$): Charged kaons can
be identified topologically via their one-prong, `kink', decays 
(e.g.\ $K\to\mu\nu$, $K\to\pi\pi^0$)~\cite{SpirosReview,DengThesis}. 
The parent kaon and charged daughter tracks are used to determine the
decay kinematics. The decay position was restricted by a fiducial cut
inside the volume of the TPC to improve the signal to background
ratio. Background sources include charged pion decays, hadronic
interactions in the TPC gas, and combinatorics. A momentum dependent
decay angle cut was used to eliminate essentially all pion decays
because they have a much smaller decay angle. 
A remaining background of the order of 15\% is corrected for
in the kaon spectra~\cite{DengThesis}.

Neutral kaons were reconstructed via their decay $\ks\to\pi^+\pi^-$.
A pair of oppositely charged tracks formed a $\ks$ decay-candidate if
their distance of closest approach to each other was less than 1~cm.
The majority of the combinatorial background was eliminated by requiring 
that the daughter tracks miss the primary vertex by at least 1.5~cm, 
and the decay vertex had to be separated from the primary vertex 
by more than 6~cm~\cite{MattThesis}. A cut on the $\mdedx$ of the daughters 
was also applied to remove a majority of the contribution from 
$\Lambda\to p\pi^-$ and $\overline{\Lambda}\to\overline{p}\pi^+$ decays.

\begin{table*}[hbt]
\caption{The mid-rapidity kaon multiplicity densities ($\dndy$) 
and $\mt$ exponential inverse slopes ($T$ in MeV) as a function of 
negative hadron multiplicity within $|\eta|<0.5$ ($\dnhm$). 
Quoted errors are uncorrelated errors (first) and 
correlated systematic errors (second). See text for details.
Systematic error on $\dnhm$ is 7\%. 
The centrality bins are as same as in Ref.~\cite{pbarPaper}.
}
\label{tab}
\begin{ruledtabular}
\begin{tabular}{c|c|cc|cc|cc}
Centrality & $dN_{h^-}$ & 
\multicolumn{2}{c|}{$K^+$} & 
\multicolumn{2}{c|}{$K^-$} & 
\multicolumn{2}{c}{$\ks$} \\ 
\cline{3-8}
bin & $\overline{\;\;\;d\eta \;\;\;}$ & 
$\dndy$ & $T$ & $\dndy$ & $T$ & $\dndy$ & $T$ \\ 
\hline
58-85\% & 17.9 & 2.46$\pm$0.07$\pm$0.32  & 241$\pm$7$\pm$19 
		& 2.32$\pm$0.06$\pm$0.30 & 238$\pm$7$\pm$19 
		& 1.82$\pm$0.04$\pm$0.27 & 253$\pm$4$\pm$15 \\
45-58\% & 47.3 	& 7.23$\pm$0.18$\pm$0.95 & 242$\pm$6$\pm$19 
		& 6.48$\pm$0.17$\pm$0.84 & 257$\pm$7$\pm$21 
		& 5.40$\pm$0.10$\pm$0.81 & 268$\pm$3$\pm$16 \\
34-45\% & 78.9 	& 11.8$\pm$0.3$\pm$1.5   & 265$\pm$6$\pm$21 
		& 10.4$\pm$0.2$\pm$1.4   & 250$\pm$6$\pm$20 
		& 9.57$\pm$0.17$\pm$1.4  & 274$\pm$3$\pm$16 \\
26-34\% & 115. 	& 17.2$\pm$0.4$\pm$2.3   & 281$\pm$7$\pm$22 
		& 15.5$\pm$0.4$\pm$2.0   & 268$\pm$7$\pm$21 
		& 14.0$\pm$0.2$\pm$2.1   & 273$\pm$3$\pm$16 \\
18-26\% & 154. 	& 23.1$\pm$0.5$\pm$3.0   & 275$\pm$7$\pm$22 
		& 20.8$\pm$0.5$\pm$2.7   & 271$\pm$7$\pm$22 
		& 18.8$\pm$0.3$\pm$2.8   & 287$\pm$3$\pm$17 \\
11-18\% & 196. 	& 28.8$\pm$0.7$\pm$3.8   & 269$\pm$6$\pm$22 
		& 26.4$\pm$0.6$\pm$3.4   & 274$\pm$7$\pm$22 
		& 23.3$\pm$0.4$\pm$3.5   & 287$\pm$3$\pm$17 \\
 6-11\% & 236. 	& 38.0$\pm$0.6$\pm$5.0   & 284$\pm$4$\pm$23 
		& 34.5$\pm$0.5$\pm$4.5   & 283$\pm$4$\pm$23 
		& 31.4$\pm$0.5$\pm$4.7   & 277$\pm$3$\pm$17 \\
  0-6\% & 290. 	& 46.2$\pm$0.6$\pm$6.1   & 277$\pm$4$\pm$22 
		& 41.9$\pm$0.6$\pm$5.4   & 277$\pm$4$\pm$22 
		& 36.7$\pm$0.6$\pm$5.5   & 285$\pm$3$\pm$17 \\
\end{tabular}
\end{ruledtabular}
\end{table*}

In all cases, the primary vertex was restricted to a limited
longitudinal range near the center of the TPC. The event centrality
was determined off-line and is based on measured charged particle
multiplicities in the TPC. A correction factor was applied to account
for losses due to limited acceptance, decay, tracking inefficiency,
and hadronic interactions. The overall efficiency ($\epsilon$),
including all these effects, was obtained from a full MC simulation
by embedding MC tracks into real events on the raw data level.
For the most central collisions, the \dedx\ method yielded
$\epsilon\simeq 20$\% at $\pt$=0.2~GeV/$c$ and 60\% at 0.7~GeV/$c$.
For the \kink\ method, $\epsilon\simeq 1.5$\% at $\pt$=1~GeV/$c$ and
0.6\% at 2~GeV/$c$. For the $\ks$ method, $\epsilon\simeq 4.5$\% at
$\pt$=1~GeV/$c$ and 7\% at 2~GeV/$c$. The efficiency increases with
decreasing event multiplicity by about 20\% of its value, from the most
central to the most peripheral bins, for the \dedx\ and \kink\
methods and by 70\% for the $\ks$ method.

Figure~\ref{spectra} shows the transverse mass spectra for the
invariant yields of $K^+$, $K^-$, and $\ks$, where 
$\mt\equiv\sqrt{\pt^2+m^2}$ and $m$ is the kaon mass. The errors shown
for the \dedx\ method are the quadratic sum of the statistical and
point-to-point systematic errors. The errors shown for \kink\ and
$\ks$ methods are statistical only. The overall systematic errors
are estimated to be 5\%, 10\%, and 10\% for \dedx, \kink, and $\ks$
methods, respectively; they are uncorrelated among the three analyses.
An additional 5\% systematic error, due to uncertainties in our MC
determination of the efficiencies, applies to all three analyses.
Figure~\ref{spectra}(c) compares $\ks$ spectra to the averaged charged
kaon spectra from the \kink\ method. As isospin asymmetry is
negligible at mid-rapidity~\cite{PhobosRatio}, the primordial $\ks$
yield is most likely equal to the average of the primordial charged
kaon yields. However, our measurements include decay products of
$\phi$ mesons which decay into charged kaons and neutral kaons with
different branching ratios. 
The effect is estimated using the measured $\phi$ spectra~\cite{phiPaper}  
to be 1-3\% between the measured charged and neutral kaons 
in the $0.2<\pt<1.0$~GeV/$c$ range.

The kaon spectra exhibit an exponential shape in $\mt$. We fit the
spectra of charged kaons (combined from \dedx\ and \kink) and $\ks$,
respectively, to an $\mt$ exponential with the inverse slope, $T$, 
and the integrated rapidity density, $\dndy$, as free parameters. 
The fit results are shown as solid lines in Fig.~\ref{spectra}. 
The fit parameters are listed in Table~\ref{tab} together with $\dnhm$, 
the negative hadron multiplicity within $|\eta|<0.5$~\cite{STARhminus}.
Systematic errors on $\dndy$ and $T$ are both about 8\% for charged
kaons, and 10\% and 6\%, respectively, for $\ks$. The systematic
errors are partially correlated between $K^+$ and $K^-$. 
An additional 5\% systematic error applies to the $\dndy$ and it is 
correlated between all three particle species.
Our charged kaon $\dndy$ results are in agreement with the recent 
PHENIX publication~\cite{PhenixPaper} 
and the point-by-point spectra agree within two standard deviation of 
systematic errors. 

\begin{figure}[hbt]
\centerline{\epsfxsize=0.5\textwidth\epsfbox[10 40 510 490]{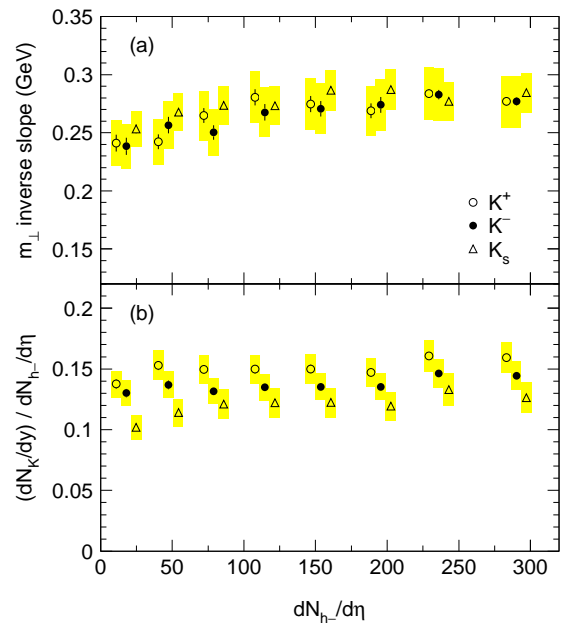}}
\caption{The centrality dependence of (a) kaon inverse slopes and 
(b) mid-rapidity kaon to negative hadron ratios. The error bars are
uncorrelated random errors; correlated systematic errors are indicated
by the shaded areas. An additional systematic error (not shown) of
5\% and 7\% applies to the $\dndy$ of kaons and $\dnhm$, respectively.
For clarity the $K^+$ and $\ks$ points are displaced in $\dnhm$.}
\label{dndy}
\end{figure}

Figure~\ref{dndy}(a) shows $T$ as a function of $\dnhm$. 
No difference is observed between the $K^+$, $K^-$ and $\ks$. 
There is an indication of a systematic increase in $T$ from $\sim$240~MeV 
in the most peripheral collisions to $\sim$280~MeV in the most central
collisions. For comparison, the kaon inverse slope is about 240~MeV
for central heavy ion collisions at the SPS 
($\snn\approx 17$~GeV)~\cite{NA49,NA44,NA44NuXu} and 
200~MeV at the AGS ($\snn\approx 5$~GeV)~\cite{E866,E917}. 
Note, however, that inverse slopes may measure a combined effect of 
thermal temperature and transverse radial flow~\cite{NA44NuXu} and 
the larger $T$ values suggest stronger radial flow at RHIC energies. 
For the most central collisions, the measured kaon $T$ is smaller than 
that of the lambda and the lambda $\dndy$~\cite{lambdaPaper} 
is about 1/3 of the kaon $\dndy$. 
As a result, the lambda yield approaches, and may even exceed 
the kaon yield for $\pt$\gsim1.5~GeV/$c$. 
A similar behaviour was observed for non-strange particles 
(pion and proton) in a similar $\pt$ region~\cite{PhenixPaper}.

Figure~\ref{dndy}(b) shows the ratio of kaon $\dndy$ to $\dnhm$ as a
function of $\dnhm$. No strong centrality dependence is observed for
the ratios, suggesting no significant change in strangeness production
mechanisms from peripheral to central collisions at this RHIC energy.
In contrast, kaon production at lower AGS and SPS energies roughly
doubles from peripheral to central collisions~\cite{E866,NA49}. 
On the other hand, the $K^+/K^-$ ratio remains constant as a function 
of centrality at all energies~\cite{E866,NA49,WangKratio}.

\begin{figure}[hbt]
\centerline{\epsfxsize=0.5\textwidth\epsfbox[20 130 510 410]{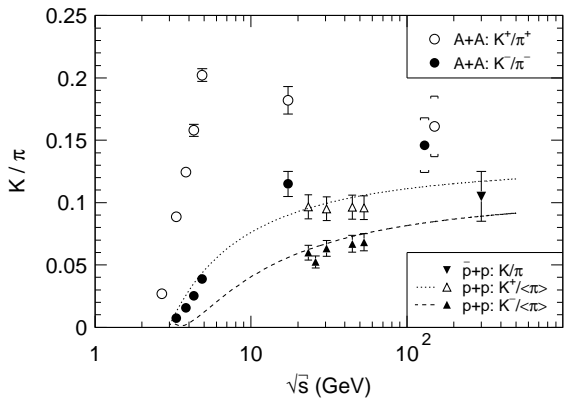}}
\caption{Mid-rapidity $K/\pi$ ratios versus $\snn$. The curves are
parameterizations to p+p data~\cite{Rossi}. The error bars show 
statistical errors. The systematic errors on the STAR data are
indicated by the caps. The STAR $K^+/\pi$ point is displaced in
$\snn$ for clarity.}
\label{roots}
\end{figure}

$K/\pi$ ratios are often used to study strangeness production
enhancement. In order to evaluate $K/\pi$, we deduce the mid-rapidity
pion $\dndypi$ in central collisions from our measurements of negative
hadrons~\cite{STARhminus}, antiprotons~\cite{pbarPaper} and $K^-$
spectra. The deduced mid-rapidity value is $\dndypi = 287 \pm 20$,
consistent with our preliminary measurement of pion spectra cited
in~\cite{HarrisQM01}. For the most central collisions, 
$K^+/\pi^- = 0.161 \pm 0.002 {\rm (stat)} \pm 0.024 {\rm (syst)}$ and 
$K^-/\pi^- = 0.146 \pm 0.002 {\rm (stat)} \pm 0.022 {\rm (syst)}$. 
The systematic errors are a quadratic sum of those on the kaon and 
the pion $\dndy$. Figure~\ref{roots} is a compilation of $K/\pi$ results 
for central heavy ion collisions. Since mid-rapidity $\pi^+/\pi^-\approx 1$ 
at RHIC~\cite{PhobosRatio}, we can readily compare our $K^+/\pi^-$
results to $K^+/\pi^+$ results from lower energies. The $K^-/\pi$
ratio steadily increases with $\snn$, while the $K^+/\pi$ ratio in
heavy ion collisions sharply increases at low energies and the maximum
value of $K^+/\pi^+$ occurs at $\snn\sim 10$~GeV. This value is
determined by the interplay between the dropping net-baryon density
with $\snn$ and an increasing $K\overline{K}$ pair production rate, as
previously noted (e.g.\ in \cite{WangKratio,WangRQMD}).

Figure~\ref{roots} also shows parameterized p+p data (curves) and data
from p+p~\cite{Rossi} and $\overline{\rm p}$+p~\cite{E735} at higher
energies. Our measurement indicates a 50\% enhancement over 
$(K^+ + K^-)/2\langle\pi\rangle$ in p+p and $\overline{\rm p}$+p collisions 
at similar energies. The enhancement in $K^-/\pi$ is similar at SPS and
RHIC, while that in $K^+/\pi$ is larger at lower energies due to the
effect of a changing net-baryon density.

In conclusion, we have reported invariant yield transverse mass
spectra and multiplicity densities of charged and neutral kaons at
mid-rapidity in Au+Au collisions at $\snn$=130~GeV at RHIC. 
The spectra are described by an exponential in transverse mass. 
The inverse slope parameters are found to increase slightly with 
collision centrality, with a value of about 280~MeV in central
collisions. No strong centrality dependence is found in the ratio of
kaon rapidity densities to negative hadron pseudo-rapidity densities.
For the most central collisions, the mid-rapidity kaon to pion ratios
are $K^+/\pi^- = 0.161 \pm 0.002 {\rm (stat)} \pm 0.024 {\rm (syst)}$
and $K^-/\pi^- = 0.146 \pm 0.002 {\rm (stat)} \pm 0.022 {\rm (syst)}$.
For central heavy ion collisions, the $K^+/\pi$ ratio is found to
increase rapidly with the collision energy and then to decrease,
while the $K^-/\pi$ ratio increases steadily. 
This behavior is consistent with the increasing pair production rate 
as the collision energy increases, and the decreasing net-baryon density 
at mid-rapidity. 
The measured $K/\pi$ ratios at RHIC show an enhancement of about 50\% 
over p+p and $\overline{\rm p}$+p collisions at similar energies.

%\acknowledgments
We wish to thank the RHIC Operations Group and the RHIC Computing
Facility at Brookhaven National Laboratory, and the National Energy
Research Scientific Computing Center at Lawrence Berkeley National
Laboratory for their support. This work was supported by the Division
of Nuclear Physics and the Division of High Energy Physics of the
Office of Science of the U.S.Department of Energy, the United States
National Science Foundation, the Bundesministerium fuer Bildung und
Forschung of Germany, the Institut National de la Physique Nucleaire
et de la Physique des Particules of France, the United Kingdom
Engineering and Physical Sciences Research Council, Fundacao de Amparo
a Pesquisa do Estado de Sao Paulo, Brazil, and the Russian Ministry of
Science and Technology.

%%%%%%%%%%%%%%%%%%%%%%%%%%%%%%%%%%%%%%%%%%%%%%%%%%%%%%%%%%%%%%%%%%%%%%

\end{document}